\documentclass[%
 reprint,
superscriptaddress,
showpacs,preprintnumbers,
 amsmath,amssymb,
 aps,
prl,
]{revtex4-1}
\usepackage{amsfonts}
\usepackage[english]{babel}
\usepackage[latin1]{inputenc}
\usepackage{color}
\usepackage{graphicx}
\usepackage{dcolumn}
\usepackage{bm}

\begin{document}

\preprint{APS/123-QED}
\title{Amplified Stimulated Terahertz Emission from Optically Pumped Graphene at Room Temperature}

\date{\today}

\author {S. Boubanga-Tombet}
\email{ stephanealbon@hotmail.com}
\affiliation{Research Institute of Electrical Communication, Tohoku University, 2-1-1 Katahira, Aoba-Ku, Sendai 980-8577, Japan}

\author{S. Chan} 
\affiliation{Nano-Japan Program, University of Pennsylvania, USA}

\author {T. Otsuji}
\affiliation{Research Institute of Electrical Communication, Tohoku University, 2-1-1 Katahira, Aoba-Ku, Sendai 980-8577, Japan}
 \affiliation{JST-CREST, Chiyoda-ku, Tokyo 1020075, Japan}
\author{A. Satou}
\affiliation{Research Institute of Electrical Communication, Tohoku University, 2-1-1 Katahira, Aoba-Ku, Sendai 980-8577, Japan}
\affiliation{JST-CREST, Chiyoda-ku, Tokyo 1020075, Japan}
\author{V. Ryzhii}
\affiliation{Computational Nano-Electronics Laboratory, University of Aizu, Japan}
\affiliation{JST-CREST, Chiyoda-ku, Tokyo 1020075, Japan}
\date{\today}

\begin{abstract}

We report on fast relaxation and relatively slow recombination dynamic of photogenerated electrons/holes in an exfoliated  graphene. Under suitable pumping the carriers dynamics can lead to non trivial feature such as negative dynamic conductivity in terahertz spectral range. Therefor we  conduct time domain spectroscopy studies and shows that graphene sheet amplifies an incoming terahertz field. The graphene emission spectra dependency on the laser pumping power shows a threshold like behavior, testifying the occurrence of the negative conductivity and the population inversion in terahertz range, paving the way for a new class room temperature graphene based terahertz lasers. 

\end{abstract}

\pacs{Valid PACS appear here}
\maketitle


Graphene is a one-atom-thick planar sheet of carbon atoms that are densely packed in a honeycomb crystal lattice\cite{NovoselovSc2004}. This material has many peculiar properties and potential applications. For example, the prediction and observation of half-integer quantum hall effect\cite{ZhangNatL2005}, finite conductivity at zero charge carrier concentration \cite{NovoselovNat2005}, perfect quantum tunneling effect \cite{KatsnelsonNatPhys2006}, ultrahigh carrier mobility  \cite{GeimNatMater2007},  including massless and gap less energy spectra. The gap less and linear energy spectra of electrons and holes lead to nontrivial features such as negative dynamic conductivity in the terahertz (THz) spectral range \cite{RyzhiiJAP2007},  which may lead to the development of a new type of THz laser \cite{DubinovAPExp2009,RyzhiiJAP2009}. This perspective generate an intense interest due to the ongoing search for viable THz detectors and emitters.

To realize such THz graphene-based devices, understanding the non-equilibrium carrier relaxation/recombination dynamics is critical.  Recently, time-resolved measurements of fast non-equilibrium carrier relaxation dynamics have been carried out for multilayer and monolayers of graphene that were epitaxially grown on SiC \cite{DawlatyAPL2008, SunPRL2008, GeorgeNanoLett2008, ChoiAPL2009, WangAPL2010} and exfoliated from highly oriented pyrolytic graphite (HOPG)\cite{KampfrathPRL2005,BreusingPRL2009}.  Several methods for observing the relaxation processes have been reported. Dawlaty et al. \cite{DawlatyAPL2008} and Sun et al. \cite{SunPRL2008} used an optical-pump/optical-probe technique and George et al. \cite{GeorgeNanoLett2008} used an optical-pump/THz-probe technique to evaluate the dynamics starting with the main contribution of carrier-carrier (cc) scattering in the first 150 fs, followed by observation of carrier-phonon (cp) scattering on the picosecond time scale. Ultrafast scattering of photoexcited carriers by optical phonons has been theoretically predicted by Ando \cite{AndoJPhySocJpn2006}, Suzuura \cite{Suzuura2008} and Rana \cite{RanaPRB2009}. Kamprath et al. \cite{KampfrathPRL2005} observed strongly coupled optical phonons in the ultrafast carrier dynamics for a duration of 500 fs by optical-pump/THz-probe spectroscopy. Wang et al. \cite{WangAPL2010} also observed ultrafast carrier relaxation by hot-optical phonons-carrier scattering for a duration of $\sim$ 500 fs  by using an optical-pump/optical-probe technique. The measured optical phonon lifetimes found in these studies were $ \sim$ 7 ps  \cite{KampfrathPRL2005}, 2-2.5 ps \cite{WangAPL2010}, and $ \sim 1$ ps \cite{GeorgeNanoLett2008}, respectively, some of which agreed fairly well with theoretical calculations by Bonini et al.\cite{BoniniPRL2007}. A recent study by Breusing et al. \cite{BreusingPRL2009} more precisely revealed ultrafast carrier dynamics with a time resolution of 10 fs for exfoliated graphene and graphite. 

In this paper we report on the fast relaxation and relatively slow recombination dynamics in optically-pumped  graphene.  The recombination process is stimulated with an THz photon probe
$\sim$ 2 ps and $\sim$ 3.5 ps after the intraband carrier relaxation has started. The observed results suggest the occurrence of negative dynamic conductivity in the THz spectral range.

\begin{figure}[t]
\includegraphics[width=9cm]{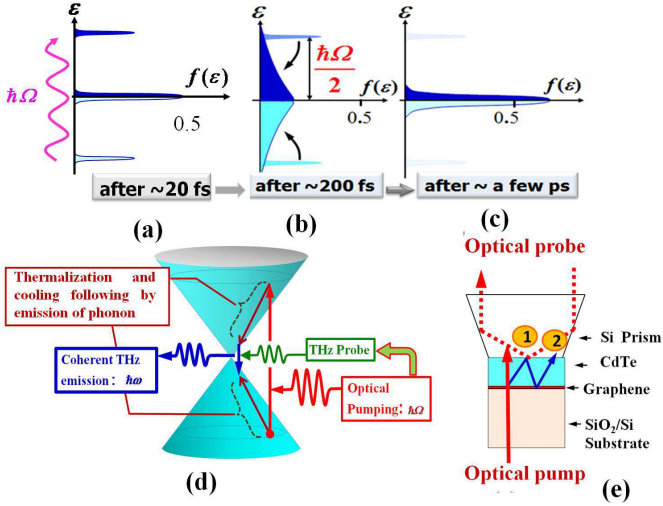}
\caption{\label{fgr:Fig1} Carrier relaxation/recombination dynamics in optically pumped graphene from few tens of fs to picoseconds after pumping (a,b and c). Measurement setup for optical-pump/THz-and-optical-probe spectroscopy. Involved process from optical pumping to emission of stimulated THz photon (d). CdTe crystal on top of the graphene sample, the optical pump and probe as well as the THz probe are represented. CdTe crystal  generates THz probe which is partially reflected at the top surface of the CdTe then subject back to the graphene. The optical probe  allows electro-optic detection of these THz electric field intensities (e).}
\end{figure}

When graphene is pumped with the infrared photon having an energy $\hbar\Omega$, electrons/holes are photogenerated via interband transitions. It has been shown that the intraband carrier  first establishes separate distributions  around the level $\varepsilon_f\pm \hbar\Omega /2$ ( $\varepsilon_f$ : Fermi energy ) within 20-30 fs after excitation (see Fig \ref{fgr:Fig1}a.).  At room temperature and/or strong pumping, collective excitations due to the cc scattering, e.g., intraband plasmons should have a strong influence on the carrier relaxation dynamics. As discussed in \cite{GeorgeNanoLett2008, KampfrathPRL2005, BreusingPRL2009} the quasi-equilibrium distributions at around $\varepsilon_f\pm \hbar\Omega /2$ is  rapidly redistributed within 200-300 fs (Fig \ref{fgr:Fig1}b.).  Then optical phonons (OPs) are emitted on the high-energy tail of the electron and hole distributions on a few picoseconds time scale(Fig \ref{fgr:Fig1}c.). This intraband relaxation process is relatively fast  and accumulates the nonequilibrium carriers around the Dirac point.  The recombination process are mainly done by interband cc  and carrier-phonons (cp) scattering.  The fast intraband relaxation (ps or less) and the interband recombination process ($\gg$ 1 ps) slowed by the density of states effects and Pauli blocking lead to the population inversion \cite{RyzhiiJAP2007}. In the case of the radiative recombination, due to the gapless symmetrical band structure, photon emissions over a wide THz frequency range are expected if the pumping photon energy is suitably chosen and the pumping intensity is sufficiently high.

In order to verify the above mentioned concept we conduct time-domain spectroscopy experiments based on an optical pump/THz-and-optical-probe technique. The time-resolved field emission properties are measured by an electro-optic (EO) sampling method in total-reflection geometry \cite{MinAPL1990}. The sample used is  exfoliated graphene on $SiO_2/Si$ substrate . The sample structure as presented in figure \ref{fgr:Fig1}e is made of one layer of Si  $<$100$>$ oriented of about 500-560 $\mu m$ of thickness and $10^{-3}$- $5*10^{-3}$ $\Omega cm$ of resistivity; following by an thermal dry layer of $SiO_2$ of 300 nm of thickness. On this substrate of about 2.5 cm x 2.5 cm size one have some islets of monolayers, bilayers and few layers of graphene. 
 To obtain the THz photon emissions from the above-mentioned carrier relaxation/recombination dynamics, the pumping photon energy (wavelength) is carefully selected to be around 0.8 eV (1550 nm), much higher than the optical phonon energy ($\sim$ 0.2 eV). The graphene sample is placed on the stage and  CdTe crystals  of about 2 mm long and 0.5 mm large are used as  THz electro-optic transceivers and placed onto the sample. A femtosecond pulsed fiber laser with full width at half-maximum of 80 fs, frequency of 20 MHz and average power of about 4 mW was used as the pumping source. More precise description of the experimental set-up can be see in Refs \cite{BoubangaAPL2010, Karasawa2010}. The laser is split  into two path used for pump and probe. The pumping laser beam linearly polarized is mechanically chopped at $\sim 1.2 KHz$ (for lock'in detection), and simultaneously focused with a beam diameter of about 40 $\mu m $ onto the sample and the CdTe from the back side, while the probing beam is cross polarized with respect to the pump beam and focused from the top side (see Fig \ref{fgr:Fig1}e). Owing to second-order nonlinear optical effects, the CdTe crystal can rectify the pump laser pulse to emit THz envelope radiation. This THz pulse is used for EO detection via pockel effect . The same THz pulse is partially reflected at the top surface of the CdTe then subject back to the graphene, working as the THz probe pulse (arrowed blue line in Fig \ref{fgr:Fig1}e) to stimulate THz photon emission via electron-hole recombination in the graphene (see Fig \ref{fgr:Fig1}d). Therefore the original data of experimental temporal response consist of the first forward propagating THz pulsation (no interaction with graphene) followed a photon echo signal (probing the graphene). The delay between these two pulsation is given by the total round-trip propagation time of the THz pulse through the CdTe. The system bandwidth is estimated to be around 6 THz, mainly limited by the Reststrahlen band of the CdTe sensor crystal.

The experiments were done with two CdTe crystal transceivers (A and B): (100) and (101) oriented respectively and of about 120 $\mu m$ and 80 $\mu m$ of thickness respectively. The graphene monolayer area investigated is of about 7000 $\mu m^2$.

\begin{figure}[h]
   \includegraphics[width=7.5cm]{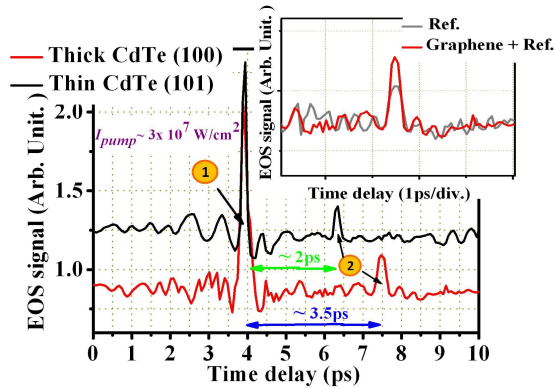}
   
   \caption{\label{fgr:Fig7}  Measured temporal responses on monolayer graphene with thicker(red line) and thiner(black line)  CdTe crystals for the pumping pulse intensity of about $3*10^7 W/cm^2$. Inset : Temporal responses of echo photon signal measured on graphene (red line) and on the area without graphene (grey line).}
\end{figure}

Figure \ref{fgr:Fig7} shows temporal responses measured on monolayer graphene with the thinner (black curve) and the Thicker crystal (red curve) for the pumping pulse intensity of about $3*10^7 W/cm^2$. These curves was plot with the same origin for comparison. One can notice that, as predicted each temporal profile is composed of two peaks from optical rectification (OR) in CdTe and the THz photon echo signal. The measured times delays between these two pulsations with crystal A (thinner) and crystal B (Thicker) of around 2 ps and 3.5 ps respectively are in a good agreement with the round trip propagation time of THz pulse through the CdTe crystal. The refractive indexes of CdTe may be obtained from \cite{Schall1999}. The OR pulse is found be broader in crystal B compare to that one of crystal A, this is the consequence of better phase-matching conditions in thin crystal. Indeed the coherent length in CdTe is estimated to be  around 100 $\mu m$  at 0.8 eV \cite{Pradarutti2008, Masaya2004}. This coherent length is maximum at 1050 mn \cite{Xu2006}. The inset of figure \ref{fgr:Fig7} present the echo signal peak measured with crystal B on graphene  as well as an reference curve (grey curve) measured on the area without graphene. One can notice that the peak obtained on graphene  is more intense than that one obtained on the substrate. This suggests that graphene amplifies the THz echo  (THz probe) signal. It is thought that this THz pulse stimulate emission from graphene amplifying the incoming echo photon by photoelectron/hole recombination in the range of the negative dynamic conductivity.

The graphene transfer function H($\omega$) is defined like  H($\omega$) = Y($\omega$)/X($\omega$) where Y($\omega$) and X($\omega$) are the Fourier transform of the second peak measured on graphene and substrate respectively.

 \begin{figure}[t]
   \includegraphics[width=8cm]{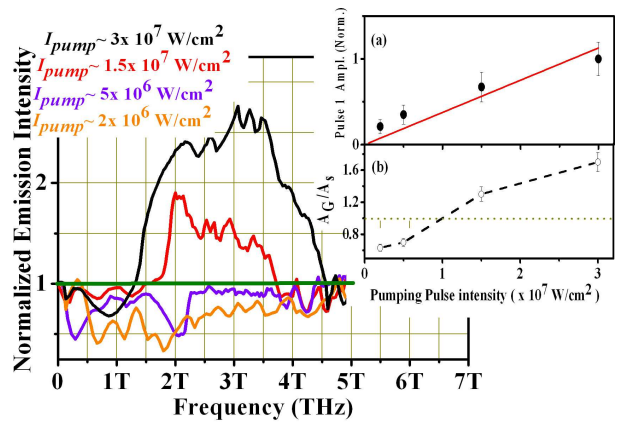}
   
   \caption{\label{fgr:Fig6} Transfer functions of monolayer graphene for different values of pumping pulse intensities. Inset : Normalized EOS signal amplitude of the first peak (upper panel) and $A_G$/$A_S$ ratio where $A_G$ and $A_S$ are the amplitude of echo signal peak measured on graphene and substrate respectively (lower panel) for different values of the pumping pulse intensities.}
\end{figure}

Figure \ref{fgr:Fig6} shows the transfer function of monolayer graphene for different values of pumping pulse intensity. One can see from these results that decreasing $I_{pump}$ drastically reduce the emission spectra and below $5*10^6 W/cm^2$ the emission completely disappear and only attenuation can be seen. We also present in the inset of figure \ref{fgr:Fig6} (lower panel) the corresponding ratio $A_G$/$A_S$ where $A_G$ and $A_S$ are the amplitude of echo signal peak measured on graphene and substrate respectively.  The graphene transfer function and $A_G$/$A_S$ shows a clear threshold like behavior testifying the occurrence of the negative conductivity and population inversion in optically pumped graphene. The threshold intensity if found to be around $10^7 W/cm^2$ . This is a good starting point for the realization of room temperature THz lasers based on graphene. The inset of figure \ref{fgr:Fig6} (upper panel) present the normalized EOS signal amplitude of the first peak ( see figure \ref{fgr:Fig7}) for different values of the pumping pulse intensities. This EOS signal is proportional to the THz electric field generated by OR in CdTe crystal. The THz emitted intensity is quadratically dependent of the infrared intensity ($I_{THz}$ $\propto$ $I^{2}_{IR}$). Since $I_{THz}$ $\propto$ $E^{2}_{THz}$, the linear depency of the EOS signal amplitude with the pumping pulse intensity is a clear evidence of its OR source.

Furthermore, to confirm the effects of the THz probe that stimulate the emission in graphene, the CdTe crystal been replaced by  a CdTe crystal having a high-reflectivity coating for IR on its bottom surface, in order to eliminate generation of the THz probe signal. In this case, no distinctive response was observed.

\begin{figure}[t]
   \includegraphics[width=7.5cm]{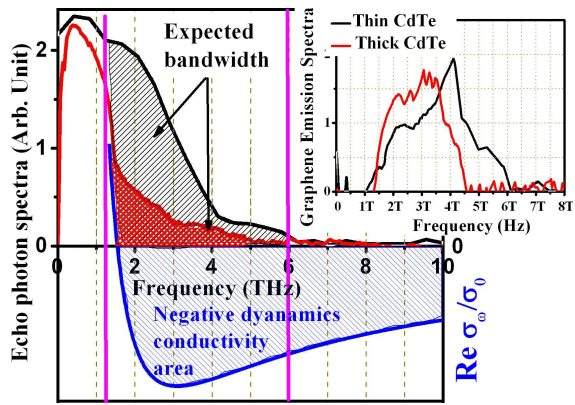}
   
   \caption{\label{fgr:Fig8} THz emission spectra of the echo photon signal measured on the area without graphene (reference) with thicker CdTe crystal (red line) and thiner CdTe (black line) crystals and the normalized dynamic conductivity for the pumping pulse intensity three time higher than the threshold pumping pulse intensity at 300 K (blue line). Inset : transfer function of monolayer graphene obtained with thicker CdTe crystal (red line) and thiner CdTe (black line) crystals.}
\end{figure}

Figure \ref{fgr:Fig8} shows the Fourier transform of the photon echo signal measured on the area without graphene (reference) with the crystal A (black curve) and the crystal B (red curve). The normalized normalized dynamic conductivity for the pumping pulse intensity three time higher than the threshold pumping pulse intensity at 300 K is also presented (see \cite{RyzhiiJAP2007, Karasawa2010}). The photon echo pulse stimulate the recombination process and the emission of THz photon in graphene within the negative dynamics conductivity area (blue shaded area). The expected graphene emission spectra bandwidth is limited at lower frequencies by the drude mechanism of THz absorption and the higher frequency limit is given by the system bandwidth which is estimated to be around 6 THz (see the horizontal solide lines in fig \ref{fgr:Fig8}). The  black and red shaded area shows the expected graphene emission bandwidth , from $\sim$ 1 THz to  $\sim$ 5 THz using the thicker crystal and from $\sim$ 1 THz to  $\sim$ 6 THz using the thinner crystal.  The broader spectra with thin crystal is due to the better phase-matching conditions.

 The inset of figure \ref{fgr:Fig8} present the transfer function of monolayer graphene obtained with crystal A (black line) and crystal B (red line). The obtained spectra is in good agreement with the above mentioned expectations. It is possible to compare these spectra within the smallest bandwidth (from $\sim$ 1 THz to  $\sim$ 5 THz). The graphene emission spectra obtained with thinner crystal is broader than that one  obtained with thicker crystal and this broadening can be understand considering the longer delay of THz photon probe in thicker crystal ($\sim$ 1.5 ps). Therefore the photocariers are  closer to the equilibrium 3.5 ps after pumping rather than 2 ps after, and the quasi-Fermi energy closer to the $\varepsilon_f$ $\approx$ 0 condition.

 Since the measurements are taken as an average, the observed response is undoubtedly a coherent process that cannot be obtained via spontaneous emission processes, providing clear evidence of stimulated emission.
The above results and discussions confirms that the THz emission from graphene is stimulated by the coherent THz probe radiation, also that the THz emission is amplified via photoelectron/hole recombination in the range of the negative dynamic conductivity.

The occurrence of population inversion in optically pumped graphene have been theoretically reported  \cite{RyzhiiJAP2007,Akira2011}. In the room temperature and/or strong pumping case, threshold pumping pulse intensity was predicted to be between $10^7 W/cm^2$ and $10^8 W/cm^2$ \cite{Akira2011}, in good agreement with our experimental results.

 In conclusion, we have successfully observed coherent amplified stimulated THz emissions arising from the fast relaxation and relatively slow recombination dynamics of photogenerated electrons/holes in an exfoliated graphene. The results provide evidence of the occurrence of negative dynamic conductivity, which can  be applied to a new type of THz laser.

This work was financially supported in part by the JST-CREST program, Japan, and a Grant-in-Aid for Basic Research (S) from the Japan Society for the Promotion of Science.


\bibliography{Manuscript}
\end{document}